\documentclass[twoside]{dis08}
\usepackage[latin1]{inputenc}
\usepackage[dvips]{graphicx,epsfig,color}
\usepackage{wrapfig,rotating}
\usepackage{amssymb,amsmath,array}

\begin{document}
\title{Multiple Interactions in {\sf Herwig++}}

\author{\underline{Manuel B\"ahr}$^1$, Stefan Gieseke$^1$ and Michael H. Seymour$^{2}$
%
\thanks{This work was supported in part by the EU Marie Curie Research
Training Network MCnet under contract MRTN-CT-2006-035606. Preprint: MCnet/08/04}
%
\vspace{.3cm}\\
%
1 - Institut f\"ur Theoretische Physik\\
Universit\"at Karlsruhe, 76128 Karlsruhe, Germany
%
\vspace{.1cm}\\
2 - School of Physics and Astronomy, University of Manchester; and\\
Physics Department, CERN, CH-1211 Geneva 23, Switzerland
}

\maketitle

\begin{abstract}
  In this contribution we describe a new model of multiple partonic
  interactions that has been implemented in {\sf Herwig++}. Tuning its
  two free parameters we find a good description of CDF underlying event
  data. We show extrapolations to the LHC and discuss intrinsic PDF
  uncertainties.
\end{abstract}

\section{Introduction}

With the advent of the Large Hadron Collider (LHC) in the near future it will
become increasingly important to gain a detailed understanding of all sources
of hadronic activity in a high energy scattering event.  An important source
of additional soft jets will be the presence of the underlying event. From the
experimental point of view, the underlying event contains all activity in a
hadronic collision that is not related to the signal particles from the hard
process, e.g.\ leptons or missing transverse energy. The additional particles
may result from the initial state radiation of additional gluons or from
additional hard (or soft) scatters that occur during the same hadron--hadron
collision. Jet measurements are particularly sensitive to the underlying event
because, although a jet's energy is dominated by the primary hard parton that
initiated it, jet algorithms inevitably gather together all other energy
deposits in its vicinity, giving an important correction to its energy and
internal structure.

In this note, based on Refs.~\cite{url:dis,Bahr:2008dy}, we want to
focus on the description of the hard component of the underlying event,
which stems from additional hard scatters within the same proton.  Not
only does this model give us a simple unitarization of the hard cross
section, it also allows to give a good description of the additional
substructure of the underlying events. It turns out that most activity
in the underlying event can be understood in terms of hard minijets. We
therefore adopt this model, based on the model \textsf{JIMMY}
\cite{Butterworth:1996zw,JimmyManual}, for our new event generator {\sf
Herwig++}{} \cite{Bahr:2008pv}. Thus far, we do not consider a
description beyond multiple hard interactions. An extension of our model
towards softer interactions along the lines suggested in
\cite{Borozan:2002fk} is planned and will also allow us to describe
minimum bias interactions. As a first step, the allowed parameter space
for such models at LHC has been identified in Ref.~\cite{Bahr:2008wk}.
 
\section{Tevatron results}

We have performed a tune of the model by calculating the total $\chi^2$
against the data from Ref.~\cite{Affolder:2001xt}. For this analysis each
event is partitioned into three parts, the \textbf{towards, away} and
\textbf{transverse} regions. These regions are equal in size in $\eta - \phi$
space and classify where particles are located in this space with respect to
the hardest jet in the event. We compare our predictions to data for the
average number of charged particles and for the scalar $p_T$ sum in each of
these regions.

\begin{wrapfigure}{r}{0.5\columnwidth}
\centerline{\includegraphics[width=0.45\columnwidth]{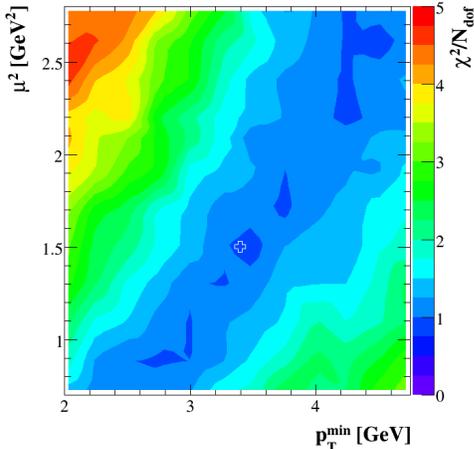}}
\caption{Contour plots for the $\chi^2$ per degree of freedom of all
    considered observables.}\label{Fig:map}
\end{wrapfigure}

The parameter space for this tune is two dimensional and consists of the $p_T$
cutoff $p_T^{\rm min}$ and the inverse hadron radius squared, $\mu^2$. In
Fig.~\ref{Fig:map} we show the $\chi^2$ contour for describing all six
observables. We have used the MRST~2001 LO\cite{Martin:2001es} PDFs built in
to {\sf Herwig++}{} for this plot, and discuss the PDF-dependence in the next
section.  For these, and all subsequent plots, we use {\sf Herwig++}\ version
2.2.1, with all parameters at their default values except the two we are
tuning and, in the next section, the PDF choice.

The description of the Tevatron data is truly satisfactory for the entire
range of considered values of $p_T^{\rm min}$. For each point on the $x$-axis
we can find a point on the $y$-axis to give a reasonable fit. Nevertheless an
optimum can be found between 3 and 4 GeV. The strong and constant correlation
between $p_T^{\rm min}$ and $\mu^2$ is due to the fact that a smaller hadron
radius will always balance against a larger $p_T$ cutoff as far as the
underlying event activity is concerned. As a default tune we use $p_T^{\rm
  min} = 3.4 \text{ GeV}$ and $\mu^2 = 1.5 \text{ GeV}^2$, which results in an
overall $\chi^2/N_{\rm dof}$ of 1.3.

\subsection{PDF uncertainties}

For precision studies it is important to quantify the extent to which hard
scattering cross sections are uncertain due to uncertainties in the PDFs. As
we have already mentioned, jet cross sections are particularly sensitive to
the amount of underlying event activity, which introduces an additional
dependence on the PDF in our model. In particular, it relies on the partonic
scattering cross sections down to small transverse momenta, which probe
momentum fractions as small as $x\sim10^{-7}$ at the LHC and $x\sim10^{-6}$ at
the Tevatron, where the PDFs are only indirectly constrained by data. One will
have measured the amount of underlying event activity at the LHC by the time
precision measurements are being made, so one might think that the size of the
underlying event correction will be known. However, in practice, jet cross
section corrections depend significantly on rare fluctuations and correlations
in the underlying event, so the correction must be represented by a model
tuned to data, rather than by a single number measured from data. This will
therefore entail in principle a retuning of the parameters of the underlying
event model for each new PDF. This would make the quantification of PDF errors
on a given jet cross section, or of extracting a new PDF set from jet data,
much more complicated than a simple reweighting of the hard scattering cross
section.

In this section we explore the extent to which this effect is important, by
studying how the predictions with fixed parameters vary as one varies the PDF.
To quantify the effect of the uncertainties within a given PDF set, we have
used the error sets provided with the CTEQ6 family, and the formula
\begin{equation*}
  \Delta X = \frac{1}{2} \ 
  \left( 
    \sum_{i=1}^{N_p} \left[ X(S_i^+) - X(S_i^-) \right]^2  
  \right)^{1/2}
\end{equation*}
from Ref.~\cite{Pumplin:2002vw}. Here, $X$ is the observable of interest and
$X(S_i^\pm)$ are the predictions for $X$ based on the PDF sets $S_i^\pm$ from
the eigenvector basis.

\begin{wrapfigure}{r}{0.5\columnwidth}
\centerline{\includegraphics[width=0.45\columnwidth]{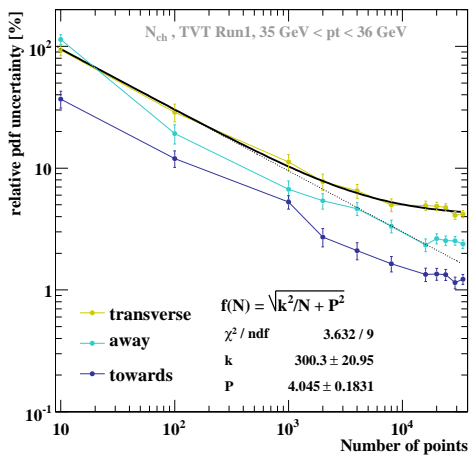}}
\caption{Relative PDF uncertainty in percent for the multiplicity
    observables. The different curves show the results for the three
    different regions defined in the experimental analysis. The PDFs
    used are CTEQ6M \cite{Pumplin:2002vw} and its corresponding error
    sets. The fit result shown as a solid line is for the transverse
    region. Also shown as a light dashed line is the fit assuming a
    purely statistical error.}
\label{Fig:pdferror}
\end{wrapfigure}

We have studied the relative PDF uncertainty, i.e.\ $\Delta X / X(S_0)$, as a
function of the number of points used for each $X(S_i^\pm)$.  We show the
result in Fig.~\ref{Fig:pdferror} for one bin corresponding to $35-36$ GeV of
the leading jet for the multiplicity observables. The final statistics are
obtained from 20M fully generated events for each PDF set and the value on the
$x$ axis is the number of events falling within this bin. We see that with
these 20M events, we have still not completely eliminated the statistical
uncertainties. However, a departure from the straight line on a log--log plot
that would be expected for pure statistical errors, $\sim1/\sqrt{N}$, is
clearly observed. We use this to extract the \emph{true} PDF uncertainty, $P$,
by fitting a curve of the form
\begin{equation*}\label{eq:fit}
  f(N) = \sqrt{\frac{k^2}{N} + P^2}
\end{equation*}
to these data. In performing the fit we get a reliable result already for a
moderate number of events. Using our fit, we have a clear indication that the
PDF uncertainty is around 4\% for the multiplicity and 4.5\% for the
$p_T^{\rm sum}$ in the transverse region.

It is note-worthy that the difference between the central values of the
MRST and CTEQ PDF sets (shown in Ref.~\cite{Bahr:2008dy}) is larger than
the uncertainty on each, at about 10 \%. Although, as we have already
mentioned, the underlying event will have already been measured before
making precision measurements or using jet cross sections to extract
PDFs, a model tuned to that underlying event measurement will have to be
used and its tuning will depend on the PDF set. We consider an
uncertainty of 5--10\% large enough to warrant further study in this
direction.

\section{LHC extrapolation}

For calculating the LHC extrapolations we left the MPI parameters at
their default values, i.e.\ the fit to Tevatron CDF data. In
Ref.~\cite{Alekhin:2005dx} a comparison of different predictions for an
analysis modelled on the CDF one discussed earlier was presented. As a
benchmark observable the charged particle multiplicity in the transverse
region was used. All expectations reached a plateau in this observable
for $p_T^{\rm ljet} > 10$~GeV. Our prediction for this observable also
reached a roughly constant plateau within this region. The height of
this plateau can be used for comparison. In Ref.~\cite{Alekhin:2005dx}
PYTHIA 6.214 \cite{Sjostrand:2001yu} ATLAS tune reached a height of
$\sim 6.5$, PYTHIA 6.214 CDF Tune~A of $\sim 5$ and PHOJET 1.12
\cite{Engel:1994vs} of $\sim 3$.  Our model reaches a height of $\sim 5$
and seems to be close to the PYTHIA 6.214 CDF tune, although our model
parameters were kept constant at their values extracted from the fit to
Tevatron data.

We have seen already in the previous section that our fit results in a flat
valley of parameter points, which all give a very good description of the
data. We will briefly estimate the spread of our LHC expectations, using only
parameter sets from this valley. The range of predictions that we deduce will
be the range that can be expected assuming no energy dependence on our main
parameters. Therefore, early measurements could shed light on the potential
energy dependence of the input parameters by simply comparing first data to
these predictions. We extracted the average value of the two transverse
observables for a given parameter set in the region $20 \text{ GeV} <
p_T^{\rm ljet} < 30 \text{ GeV}$. We did that for the best fit points at three
different values for $p_T^{\rm min}$, namely 2 GeV, 3.4 GeV and 4.5 GeV.

\begin{center}
  \begin{tabular}{|l|c|c|}
    \hline
    LHC predictions &$\langle N_{\rm chg}\rangle^{\rm transv}$ & $\langle
    p_T^{\rm sum}\rangle^{\rm transv} [\text{GeV}]$\\
    \hline
    TVT best fit& $5.1 \pm 0.3$ & $5.0 \pm 0.5$\\
    \hline
  \end{tabular}
\end{center}



\begin{footnotesize}
\bibliographystyle{unsrt} \bibliography{lit}
\end{footnotesize}

\end{document}